\documentclass[prb,twocolumn,showpacs,amssymb]{revtex4}

\usepackage{graphicx}
\usepackage{ifthen}
\usepackage{ifpdf}
\usepackage{subfigure}

\usepackage{latexsym}
\usepackage{amssymb}
\usepackage{bm}
\newcommand{\half}{\mbox{\small $\frac{1}{2}$}}
\newcommand{\pd}[2]{\frac{\partial #1}{\partial #2}}

\newcommand{\im}{\mbox{Im}}

\newcommand{\eexp}{\mbox{e}^}

\newcommand{\kk}{{\bf k}}
\newcommand{\rr}{{\bf r}}

\newcommand{\uu}{{\bf u}}

\newcommand{\hh}{{\bf h}}

\newcommand{\mm}{{\bf m}}
\newcommand{\text}[1]{\mbox{#1}}

\newcommand{\A}{{\bf A}}
\newcommand{\dudzz}{\frac{\partial \uu}{\partial z'}}

\newcommand{\hatz}{\hat{z}}
\newcommand{\dudz}{\frac{\partial \uu}{\partial z}}
\newcommand{\beq}[1]{\begin{eqnarray}\ifthenelse{#1=-1}{\nonumber}
{\ifthenelse{#1=0}{}{\label{e#1}}}}
\newcommand{\eeq}{\end{eqnarray}}
\newcommand{\hide}[1]{}
\ifpdf
\def\ext{pdf}
\else
\def\ext{eps}
\fi

\begin{document}

\title{Single vortex fluctuations in a superconducting chip as generating dephasing and spin flips in cold atom traps}

\author{Amir Fruchtman and Baruch Horovitz }

\affiliation{Department of Physics, Ben
Gurion University of the Negev, Beer Sheva 84105, Israel}


 \begin{abstract}
 We study trapping of a cold atom by a single vortex line in an extreme type II superconducting chip, allowing for pinning and friction.
We evaluate the atom's spin flip rate and its dephasing due to the vortex fluctuations in equilibrium and find that they decay rapidly when the distance to the vortex exceeds the magnetic penetration length.
We find that there are special spin orientations, depending on the spin location relative to the vortex, at which spin dephasing is considerably reduced while perpendicular directions have a reduced spin flip rate. We also show that the vortex must be perpendicular to the surface for a general shape vortex.
\end{abstract}


\maketitle

\section{Introduction}
A significant goal of atom chip experiments is to trap cold atoms near a surface at the submicron scale. The magnetic fluctuations near a metallic surface lead to significant spin flip (sf) transition towards untrapped magnetic sublevels and hence loss of atoms from the trap \cite{emmert}. This has led to theoretical study of superconducting atom chips predicting a significant reduction of noise \cite{skagerstam1,hohenester,skagerstam2} of 6-12 orders of magnitude. The reduction is more significant when the atom's distance $z$ from the surface is in the range $\lambda<z<\delta$ where $\lambda$ is the London penetration length and $\delta$ is the skin depth of the normal phase; e.g. for Nb chip and Rb atom with sf frequency of $\nu=560$KHz we have $\lambda=35$nm and $\delta=150\mu$m. Experimental data \cite{hufnagel,kasch}  have reached $\approx 30\mu$m showing, however, an enhancement of the lifetime by a factor $\approx 10$.

In current atom-chip experiments, DC magnetic fields of the order of $10-100$G are applied orthogonally to
slabs of type-II superconductors, resulting in vortices within the superconducting material. Dynamics of high density vortices were considered as a source of noise \cite{scheel}, and the relation to flux flow was studied \cite{nogues}. Furthermore, magnetic fluctuations lead to dephasing of coherent spin states \cite{scheel,folman}.

Further interest in vortices is their control of the magnetic field close to the surface hence producing a magnetic trap. Stable traps due to vortices on a thin superconducting disc were demonstrated experimentally \cite{shimizu}. Other shapes of chip can lead to programmable magnetic trap geometries \cite{muller1}. Furthermore, isolated vortices can be generated in a remanent state, leading to stable traps \cite{muller2}. Near-field noise is expected to be reduced due to the proximity of the superconductor, and technical noise is minimized as no transport current is needed to create the trap.

In the present work we consider fluctuations of a single vortex and the resulting dephasing and sf rate of an atom above the surface. We consider the limit of type II superconductors, i.e. the ratio $\kappa=\lambda/\xi$ is large, where $\xi$ is the coherence length. We also assume that dissipation is dominated by the vortex frictional motion, i.e. the surrounding superconductor is non-dissipative; the latter has been studied separately \cite{skagerstam1,hohenester,skagerstam2}.
We consider the vortex displacement $\uu(z)$, where $z<0$ is the distance from the surface at $z=0$, and show that it must end perpendicular to the surface for large $\kappa$, i.e. $d\uu/dz|_0=0$. While this boundary condition has been previously used \cite{placais,sonin} it was not explicit in related works \cite{brandt,carneiro}, and in fact configurations which deviate from this condition were considered \cite{carneiro}. In addition to elastic and external forces, the vortex responds to pinning and friction forces. We consider strong pinning due to a columnar defect, or weak pinning where friction dominates.
 We find that the magnetic fluctuations decay rapidly at distances beyond $\lambda$, yet even at short distance they have significant minima corresponding to a location dependent eigenvector. Choosing the trap direction along this eigenvector reduces dephasing \cite{scheel,folman} considerably, while choosing it in the perpendicular direction reduces the sf rate.

The strategy is to consider a magnetic dipole $\mm$ at position $\rr=0,z_0>0$ ($\rr$ and $\uu(z)$ are 2-dimensional vectors) that emits a magnetic field $\hh^i(\rr,z)$ with frequency $\omega$ that is incident on the surface $z=0$. The boundary conditions determine a reflected magnetic field $\hh^r(\rr,z)$, hence a response function $h_i^r(0,z_0)=\alpha_{i,j}m_j$. Consider a spin polarized in a direction ${\hat {\bf n}}$ which is the static magnetic field at the trap center.
The fluctuation dissipation theorem determines then the magnetic fluctuations, and the sf rate of an atom with moment $\mu$ is given by the Golden rule as \cite{agarwal,henkel}
\beq{01}
\Gamma_{sf}=\frac{2\mu^2}{\hbar^2}\frac{\sum_i Im[\alpha_{ii}]}{\eexp{\hbar\omega /k_BT}-1}\,.
\eeq
Here $\omega$ is the transition frequency between the spin levels, $T$ is the temperature and $i$ sums on the two perpendicular directions to ${\hat {\bf n}}$ (for spin $>1$ one needs 2 or more transitions of the form (\ref{e01})).
In contrast, dephasing is caused by the fluctuations in the energy difference of two trapped magnetic
sublevels \cite{scheel,folman}, hence it is determined by $Im[\alpha_{ii}]$ where $i$ in now in the ${\hat n}$ direction.

 The evaluation of the reflected wave proceeds in the following steps: (i) evaluate the magnetic field at $z<0$ and then $\hh^r(\rr,z)$ in term of a general vortex shape $\uu(z)$, (ii) find an equation of motion for $\uu(z)$, including elastic, external source, friction and pinning forces, and solve as a response to $\mm$, (iii) combine (i) and (ii) to find the response $\alpha_{ij}$ and hence the dephasing and sf rates.

\section{Magnetic fields}
Consider a superconductor occupying half space at $z<0$ with a single vortex line whose equilibrium position is at $\rr=\uu_0$, i.e. it is  perpendicular to the surface. Allowing for fluctuations, the vortex position becomes $\uu(z)$ so that the local vortex orientation is ${\hat z}+d\uu/dz$ and the superconducting phase $\varphi(\rr,z)$ satisfies
\beq{02}
\nabla \times \nabla \varphi = 2\pi\delta^2(\rr-\uu)(\hat{z} + \pd{\uu}{z})
\eeq
This equation satisfies $\oint_\gamma \nabla\varphi\cdot\vec{dl} = 2\pi $ with the line integral in the $\rr$ plane around $\rr=\uu(z)$ at any given $z<0$. It is assumed that $\uu(z)$ is single valued, i.e. the vortex does not bend by more than $\pi/2$. Eq. (\ref{e02}) can be solved as
\beq{03}
\nabla \varphi = \frac{(\hat{z}+\pd{\uu}{z})\times(\rr-\uu)}{(\rr-\uu)^2}
\eeq
The vector potential $\A(\rr,z)$ is a solution of London's equation \cite{tinkham}
\beq{04}
\A + \lambda^2 {\bm\nabla}\times{\bm\nabla}\times \A = \frac{\phi_0}{2\pi}{\bm\nabla}\varphi
\eeq
where $\phi_0$ is the flux quantum. The equation for the magnetic field $\hh(\rr,z)={\bm\nabla}\times\A(\rr,z)$, or its Fourier transform $\hh(\rr,z)=\int_k \hh_k(z)e^{i\kk\cdot\rr}$ where $\int_k\equiv \int d^2k/(2\pi)^2$, is
\beq{05}
[\frac{1}{\lambda^2}+k^2-\pd{^2}{z^2}]\hh_k(z) = \frac{\phi_0}{\lambda^2}e^{-i\kk\cdot\uu(z)}(\hatz+\dudz)
\eeq
The solution is found by the Green's function $G_k(z,z')$ for $\bar\hh_k(z) = \hh_k(z) - \hh_k(0)e^{\rho z}$, with $\rho = \sqrt{\frac{1}{\lambda^2}+k^2}$, which has the desirable boundary condition $\bar\hh_k(0) = 0$,
\beq{06}
G_k(z,z') = \frac{1}{2\rho}[\eexp{-\rho|z-z'|}-\eexp{\rho(z+z')}]
\eeq
The form of this Green's function reproduces the effects of image vortices \cite{brandt,carneiro} and allows for straightforward calculations for a general shape vortex. Corrections to (\ref{e06}) and to the following results are of order $O(\eexp{-L/\lambda})$ for a finite thickness $L$ of the superconductor.

It is convenient to shift the particular solution of (\ref{e05}) ${\bar \hh}_p(\rr,z)$ to  $\hh_p = \bar{\hh}_p-\lambda^2\nabla(\nabla\cdot\bar{\hh}_p)$ so that ${\bm\nabla}\cdot\hh_p=0$, hence
\beq{07}
\hh_p &=& \frac{\phi_0}{\lambda^2}\int_k\int_{z'} e^{i\kk\cdot[\rr-\uu(z')]}[G_k(z,z')(\hatz+\dudzz)\nonumber\\
&+&\lambda^2e^{\rho(z+z')}(i\kk+\rho\hatz)]
\eeq
The overall solution then is $\hh(\rr,z)=\hh_p(\rr,z)+\hh_0(\rr,z)$ where $\hh_0(\rr,z)$ is a solution of the homogenous part of (\ref{e04}), to be determined by matching with the external fields.

We proceed now to study boundary conditions.
 For low frequencies $\omega$ the dominant term in the dynamics is friction, linear in $\omega$. The Maxwell equation in the vacuum $z>0$ is then $\nabla\times\nabla\times \A =0$, neglecting the $\frac{\omega^2}{c^2}\A$ term. The current vanishes at the surface, i.e. $({\bm\nabla}\times\hh)_z(\rr,0)=0$,
and as usual, $\hh(\rr,z)$ is continuous across $z=0$. Note that ${\bm\nabla}\cdot{\bf A}=\frac{\phi_0}{2\pi}\nabla^2\varphi\theta(-z)$ leads to a jump in $\partial_zA_z$, however this boundary condition is not needed for $\hh(\rr,z)$ (Eq. \ref{e10} below) within our quasistatic limit.

The vector potential at $z>0$ has incoming and reflected components
\beq{08}
\A(\rr,z>0) = \int_k \A^i_k(z) \eexp{i\kk\cdot\rr+k z}+\int_k \A^{r}_k(z) \eexp{i\kk\cdot\rr-k z}
\eeq
where $k=|\kk|$, the gauge is $(i\kk+k \hatz)\cdot\A^i = (i\kk-k \hatz)\cdot\A^r = 0$ and $\A_i = \frac{2\pi}{c}e^{-k z_0}(i\hat{k}+\hat{z})\times\mm$ is the dipole's radiation.

The continuity of magnetic fields can be written as
\beq{09}
(i\kk+k \hatz)\times\A^i + (i\kk-k \hatz)\times\A^r = \hh_p^k(0)+\hh_0^k.
\eeq
where $\hh_p(\rr,z)=\int_k\eexp{i\kk\cdot\rr}\hh_p(\kk,z),\, \hh_0(\rr,z)=\int_k \eexp{i\kk\cdot\rr+\rho z}\hh_0^k$. Applying $(i\kk-k\hatz)$ on (\ref{e09}) eliminates $\A^r$ and with
$({\bm\nabla}\times\hh)_z(\rr,0)=0$ and using $({\bm\nabla}\times\hh_p)_z(\rr,0)=0$ the total field can be written as
\beq{10}
&\hh(\rr,z) &= \phi_0\int_k\int_{z'} e^{i\kk\cdot[\rr-\uu(z')]}[\frac{1}{\lambda^2}G_k(z,z')(\hatz+\dudzz)\nonumber\\
&+&
\eexp{\rho(z+z')}(\rho-k)(\hatz-i\hat{k})]\nonumber\\
&-&\int_k \eexp{i\kk\cdot r+\rho z}\frac{2}{\rho+k}[(i\kk\times\hatz)\cdot\A^i](k\hatz+i\rho\hat{k}).
\eeq
Applying $(i\kk+\rho\hatz)$ on (\ref{e09}) identifies, after some algebra, our goal
\beq{11}
\hh^r(\rr,z) = [S]-\frac{\phi_0}{\lambda^2}\int_k\int_{z'}\eexp{i\kk\cdot[\rr-\uu(z')]}\eexp{-k z + \rho z'}\frac{i\hat{k}-\hatz}{k+\rho}
\eeq
where $[S] =  \int_k e^{i\kk\cdot\rr-v z}\left(\frac{v-\rho}{v+\rho}(i\kk\times\hatz)\cdot\A^i\right)[i\hat{k}-\hatz]$ stands for a pure superconductor response, i.e in the absence of a vortex; this term does not contribute to the magnetic fluctuations.

\def\imgnoisesize{0.75}
\begin{figure*}
\parbox{\textwidth}{\centering
{\includegraphics[trim = 120px 40px 35px 50px, clip,width=\imgnoisesize\columnwidth]{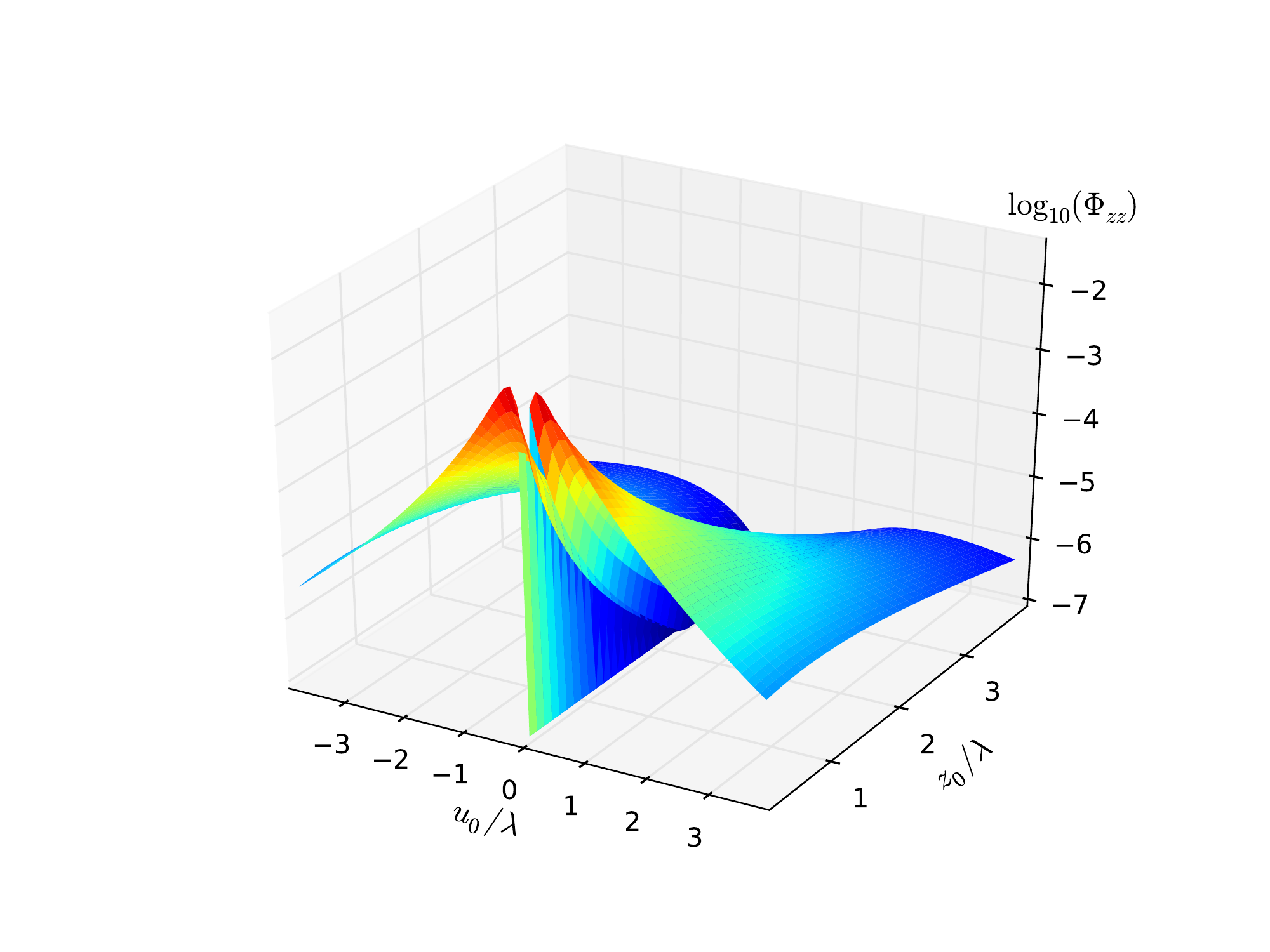}}
{\includegraphics[trim = 100px 40px 55px 50px, clip,width=\imgnoisesize\columnwidth]{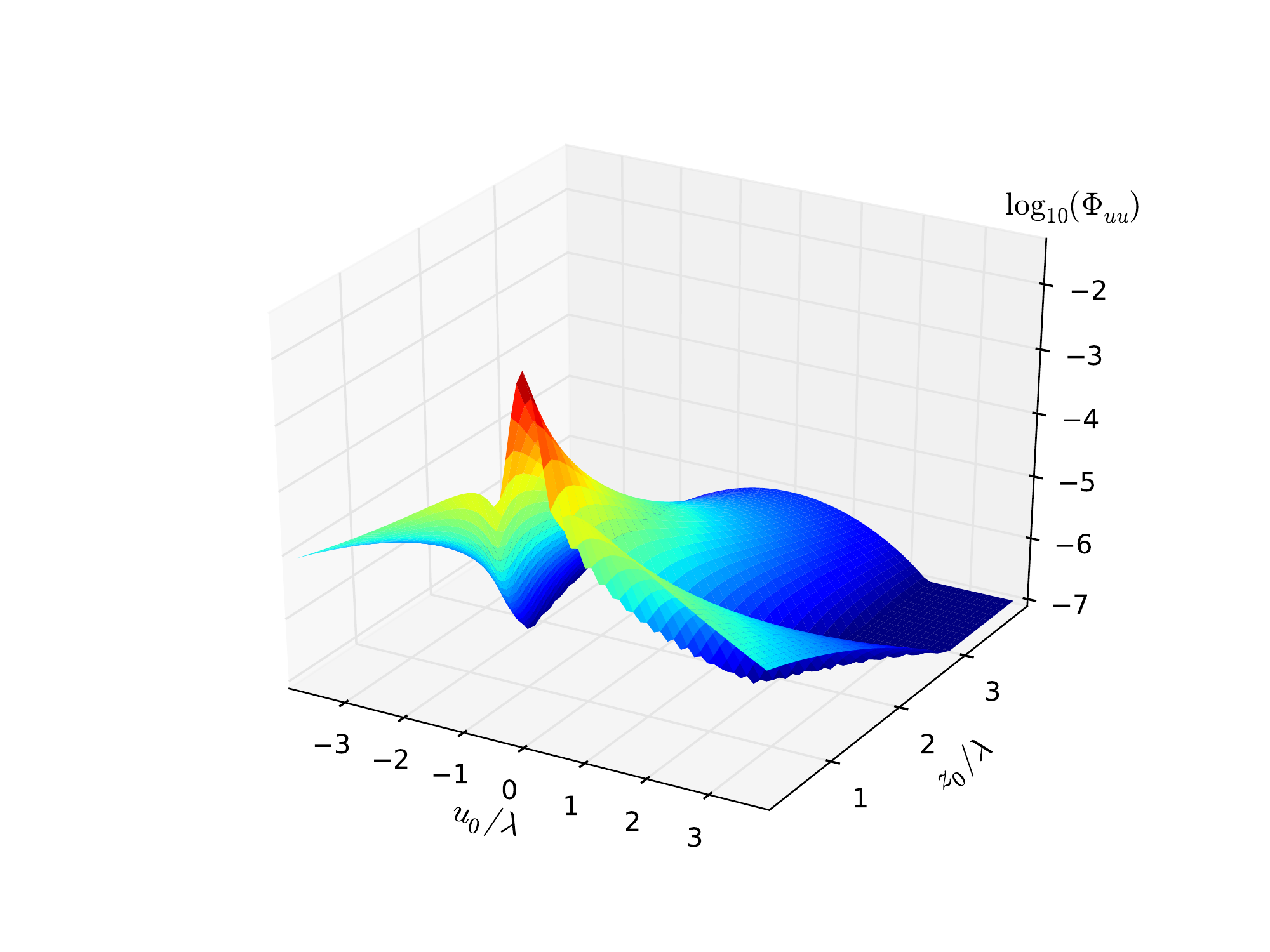}}
}
\parbox{\textwidth}{\centering
{\includegraphics[trim = 120px 40px 35px 50px, clip,width=\imgnoisesize\columnwidth]{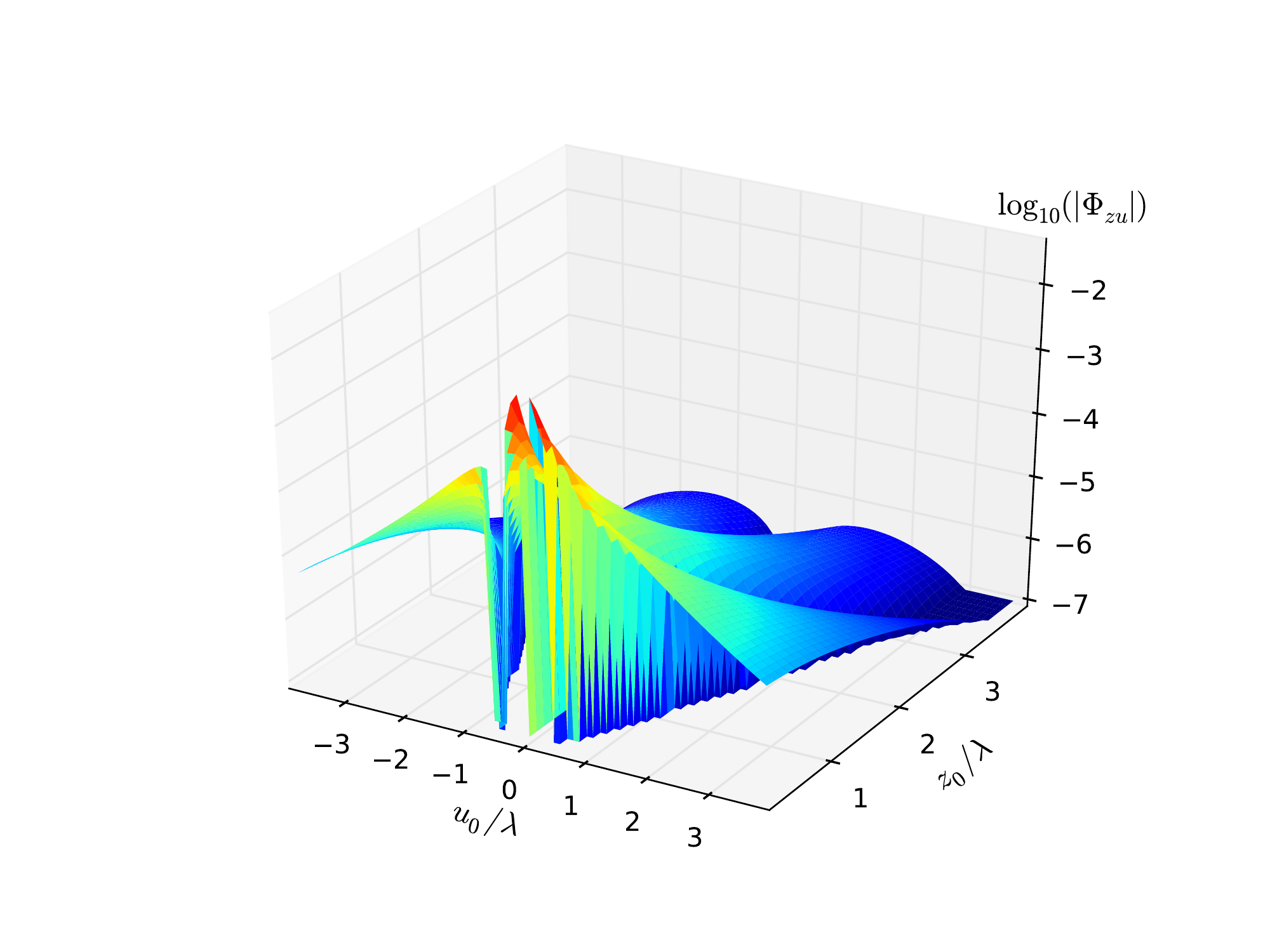}}
{\includegraphics[trim = 100px 40px 55px 50px, clip,width=\imgnoisesize\columnwidth]{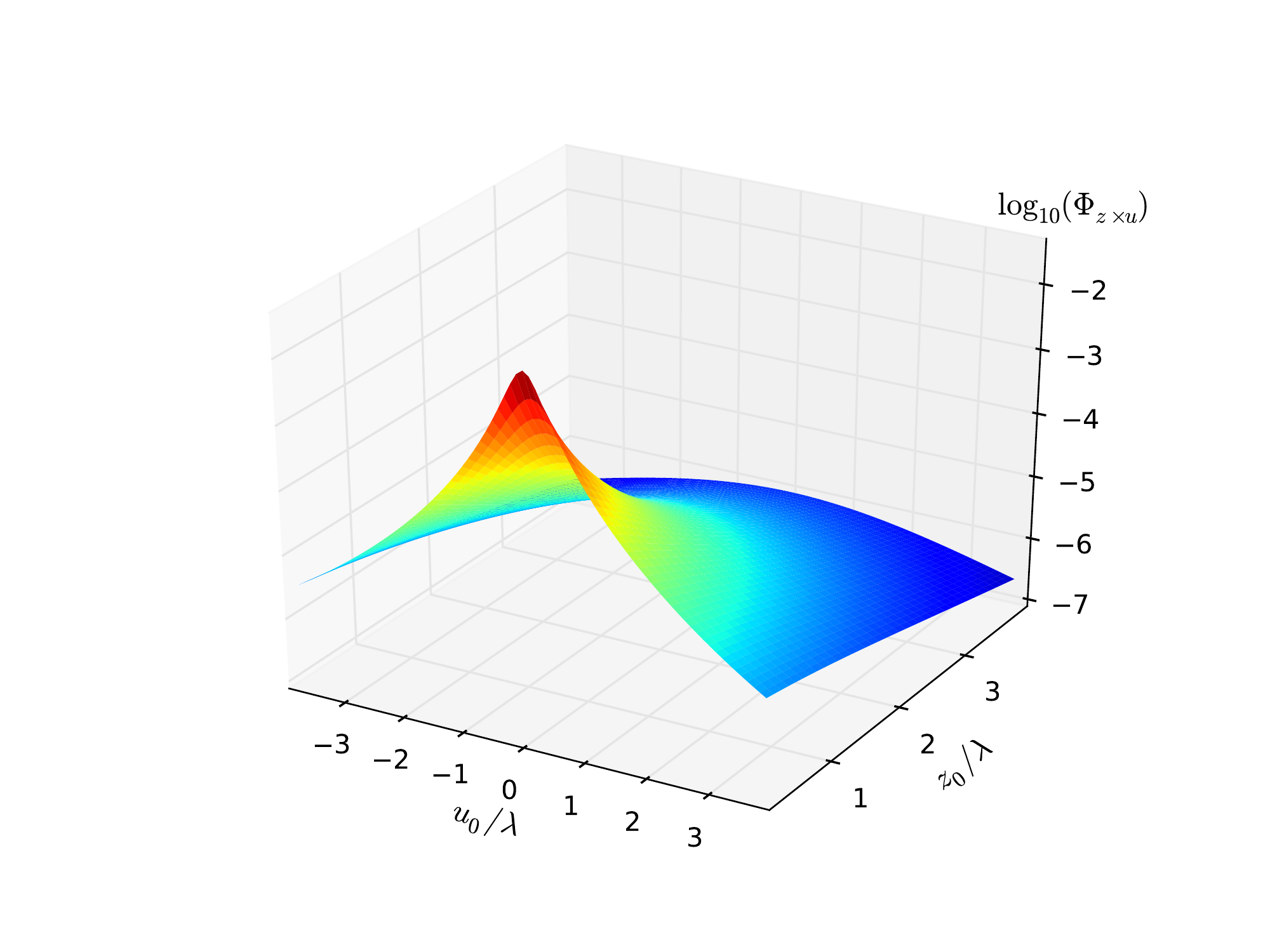}}
}
\caption{Noise components as function of $z_0,u_0$, plotted for $z_0/\lambda>0.1$. Parameters are $\bar\alpha=0.1,\,
\kappa=100$. Note that $\Phi_{zu}$ changes sign and $\Phi_{uu}$ has a minimum, both at $z_0^2\approx 2u_0^2$, while $\Phi_{zz}(0,u_0)=\phi_{zu}(0,u_0)=0$.}
\end{figure*}

\section{Vortex equation of motion}

We derive here the forces on the vortex: an elastic force, a Lorentz force due the source field, and add friction and pinning forces. The intrinsic forces can be derived from a London free energy in terms of ${\bm\nabla}\varphi$ (Eq. \ref{e03}) as a variation on $\uu(z)$. The result is the familiar Lorenz force
\beq{12}
{\bf F}_{\uu}(z)&=&\frac{\phi_0}{4\pi}(\nabla\times\hh)\times(\hat{z}+\pd{\uu}{z})_{||}\nonumber\\
&=&{\bf F}_{\uu}^{elas}(z)+{\bf F}_{\uu}^{source}(z)
\eeq
where $||$ denotes the $x,y$ components and $(\nabla\times\hh)$ is evaluated at $\rr=\uu(z)$. Here ${\bf F}_{\uu}^{source}(z)$ is
proportional to the source $\A^i$ while ${\bf F}_{\uu}^{elas}(z)$ is the intrinsic elastic force which is independent of $\A^i$. ${\bf F}_{\uu}^{elas}(z)$ is found from the 1st term of (\ref{e10}) and is evaluated by linearizing in $\uu(z')-\uu_0$, Taylor expanding $\uu(z)-\uu(z')$, integrating $z'$ and keeping only terms that diverge as $\xi\rightarrow 0$,
\beq{13}
{\bf F}_{\uu}^{elas}(z) \rightarrow \frac{\phi_0^2}{(4\pi\lambda)^2}\left\{-\dudz\int\frac{k dk}{\rho}\eexp{\rho z}+\pd{^2\uu}{z^2}\int \frac{k dk}{\rho^2}\right\}
\eeq
The first term diverges at $z=0$ as $1/\xi$ which is the upper limit on the $k$ integration. The use of this upper limit is a qualitative description and is valid only for $\ln\xi$ terms, where a change in the $\xi$ coefficient is a relatively small correction. To be consistent with the starting Eq. (\ref{e02}) we must eliminate the large $1/\xi$ force and impose a boundary condition
\beq{14}
\dudz|_{z=0}=0\,.
\eeq

We note that that the sign of this force pushes the vortex towards (\ref{e14}), i.e. the latter is stable.
This boundary condition has the intuitive interpretation that the current $\sim \frac{\phi_0}{2\pi}{\bm\nabla}\phi-{\bf A}$ (Eq. \ref{e04}) near the vortex core is dominated by ${\bm\nabla}\phi$, which in turn is perpendicular to the vortex direction (Eq. \ref{e03}). Since this current must be parallel to the surface at $z=0$ Eq. (\ref{e14}) follows. In fact this boundary condition has been used previously in context of He II \cite{placais} and superconductors \cite{sonin}. Our derivation of (\ref{e14}) is rigorous (for large $\kappa$) as it includes the full ${\bf A}$ and reflections from the surface.
For $\kappa$ that is not large one has to go beyond the phase only description of Eq. (\ref{e02}) and use coupled phase - amplitude equations \cite{tinkham}. The 2nd term of (\ref{e13}) is $\sim\ln\kappa$ and in the following we take it as the dominant term in the elastic force. The result is then the well known elastic coefficient for bending of the flux line due to change in its length \cite{tinkham}
\beq{15}
{\bf F}_{\uu}^{elas}(z)=\frac{\phi_0^2}{(4\pi\lambda)^2}\ln\kappa \pd{^2\uu}{z^2}\,.
\eeq

The external source contribution to the force is found from the 2nd term of (\ref{e10}) and in terms of the source $\mm$ it becomes
\beq{16}
{\bf F}_{\uu}^{source}(z)= \frac{\phi_0}{\lambda^2 c}\int_k e^{i\kk\cdot\uu_0+\rho z - k z_0}\frac{i\kk}{\rho+k}[(i\hat{k}+\hatz)\cdot\mm)].\nonumber\\
\eeq

 The dynamics are dominated by a friction term $\eta\frac{\partial u}{\partial t}$ where $\eta$ can be estimated from the Bardeen-Stephen result \cite{bs} (BS), $\eta=\phi_0^2/(2\pi\xi^2c^2\rho_n)$ where $\rho_n$ is the resistivity of the normal state. Finally we add a pinning term $\alpha [\uu(z)-\uu_0]$ that attempts to fix the vortex at the location $\uu_0$, i.e. a columnar defect. For this strong pinning case one can estimate $\alpha$ as the condensation energy density, hence defining
 $\alpha=\half{\bar\alpha} \phi_0^2/(4\pi\xi\lambda)^2$ strong pinning has ${\bar\alpha}\approx 1$. The equation of motion is then
 \beq{17}
 \frac{\phi_0^2}{(4\pi\lambda)^2}\ln\kappa\pd{^2\uu}{z^2}+i\omega\eta\uu-\alpha(\uu-\uu_0) = -{\bf F}_{\uu}^{source}
 \eeq
The boundary conditions are (\ref{e14}) and $\uu\rightarrow \uu_0$ at $z\rightarrow -\infty$.
The solution is, defining $ {\bf F}_{\uu}^{source}(z)= \int_k F_k e^{\rho z}$,
\beq{18}
\uu(z)-\uu_0 = \frac{-(4\pi\lambda)^2}{\phi_0^2\ln\kappa}\int_k \frac{F_k}{\rho^2-1/l^2}\left(\eexp{\rho z}-\rho l \eexp{z/l}\right)
\eeq
where $\frac{1}{l^2} = \frac{\alpha-i\omega\eta}{\phi_0^2\ln\kappa}(4\pi\lambda)^2$.

\section{response}
The response is identified from Eq. (\ref{e11}), expansion in $\uu(z)-\uu_0$ and the solution (\ref{e18}).
The angular integrations in Eqs. (\ref{e11},\ref{e18}) can be done analytically, leading to a double integral on $k,k'$ that is evaluated numerically.
The relative significance of the friction and pinning is controlled by the ratio
\beq{19}
\frac{\omega\eta}{\alpha}=\frac{16\pi\lambda^2\omega}{c^2{\bar\alpha}\rho_n}
\eeq
where the BS friction \cite{bs} is used.
For typical parameters of magnetic traps and type II superconductors we use $\lambda\approx 100$nm, $\omega\approx 1$MHz and $\rho_n\omega\approx 10^{-12}$, hence we estimate the ratio (\ref{e19}) as $10^{-5}/{\bar\alpha}$. Therefore, unless pinning is very weak we can expand in this ratio, leading to a response linear in $\omega\eta$. The response has then the form
\beq{20}
\im\alpha^{(1)}_{ij} &&= \frac{4\omega\eta}{\lambda^3 \alpha c\ln\kappa}\times \\
&&\left(   \begin{array}{ccc}
\Phi_{zz}(u_0,z_0)&\Phi_{zu}(u_0,z_0)&0\\
\Phi_{uz}(u_0,z_0)&\Phi_{uu}(u_0,z_0)&0\\
0&0&\Phi_{z\times u}(u_0,z_0)
\end{array}\right)\nonumber
\eeq
where the cartesian axes are chosen in the directions of $\hatz,\uu_0, \hatz\times\uu_0$. We note that there are two finite off-diagonal terms  $\Phi_{uz}(u_0,z_0)=\Phi_{zu}(u_0,z_0)$, that are equal as required by Onsager's reciprocity theorem.
We show the 4 components of the noise $\Phi_{ij}(u_0,z_0)$ in fig. 1. We note that an $m_z$ dipole placed above a vortex (i.e. $u_0=0$) does not produce a force on the vortex hence $\Phi_{zz}(0,z_0)=0$ (vanishing as $\sim u_0^2$) and $\Phi_{zu}(0,z_0)=0$ (vanishing as $\sim |u_0|$). Furthermore, $\Phi_{uu}(u_0,z_0)$ has a minimum and $\Phi_{zu}(u_0,z_0)$ changes sign, both at $z_0^2= 2u_0^2$ at large $z_0, u_0$ while for small $z_0$ this continues at a somewhat smaller $z_0/u_0$ ratio. For $z_0^2+u_0^2 >> \lambda^2$ we find
\beq{21}
\Phi_{zz} &=& f(\alpha)\frac{9(z_0u_0)^2}{(z_0^2+u_0^2)^5} \nonumber\\
\Phi_{uu} &=&  f(\alpha)\frac{(z_0^2-2u_0^2)^2}{(z_0^2+u_0^2)^5} \nonumber\\
\Phi_{zu} &=& f(\alpha) \frac{3z_0u_0(z_0^2-2u_0^2)}{(z_0^2+u_0^2)^5} \nonumber\\
\Phi_{z\times u} &=& f(\alpha)\frac{1}{(z_0^2+u_0^2)^3} \nonumber\\
f(\alpha)&=& \frac{\lambda^6}{2(1+\sqrt{\tilde\alpha})^4}[\frac{1+4\tilde\alpha}
{\sqrt{\tilde\alpha}}+4+\tilde\alpha]
\eeq
where $\tilde\alpha={\bar\alpha}\kappa^2/\ln\kappa$.
Hence all $\Phi_{ij}(u_0,z_0)$ decay rapidly beyond $\lambda$, with asymptotic forms
$\Phi_{ij}(u_0,z_0)\sim z_0^{-\sigma}$, with $\sigma=8,7,6,6$ for the $zz,zu,uu$ and $z\times u$ components, respectively.

\begin{figure}[htb]
\includegraphics[scale=0.38]{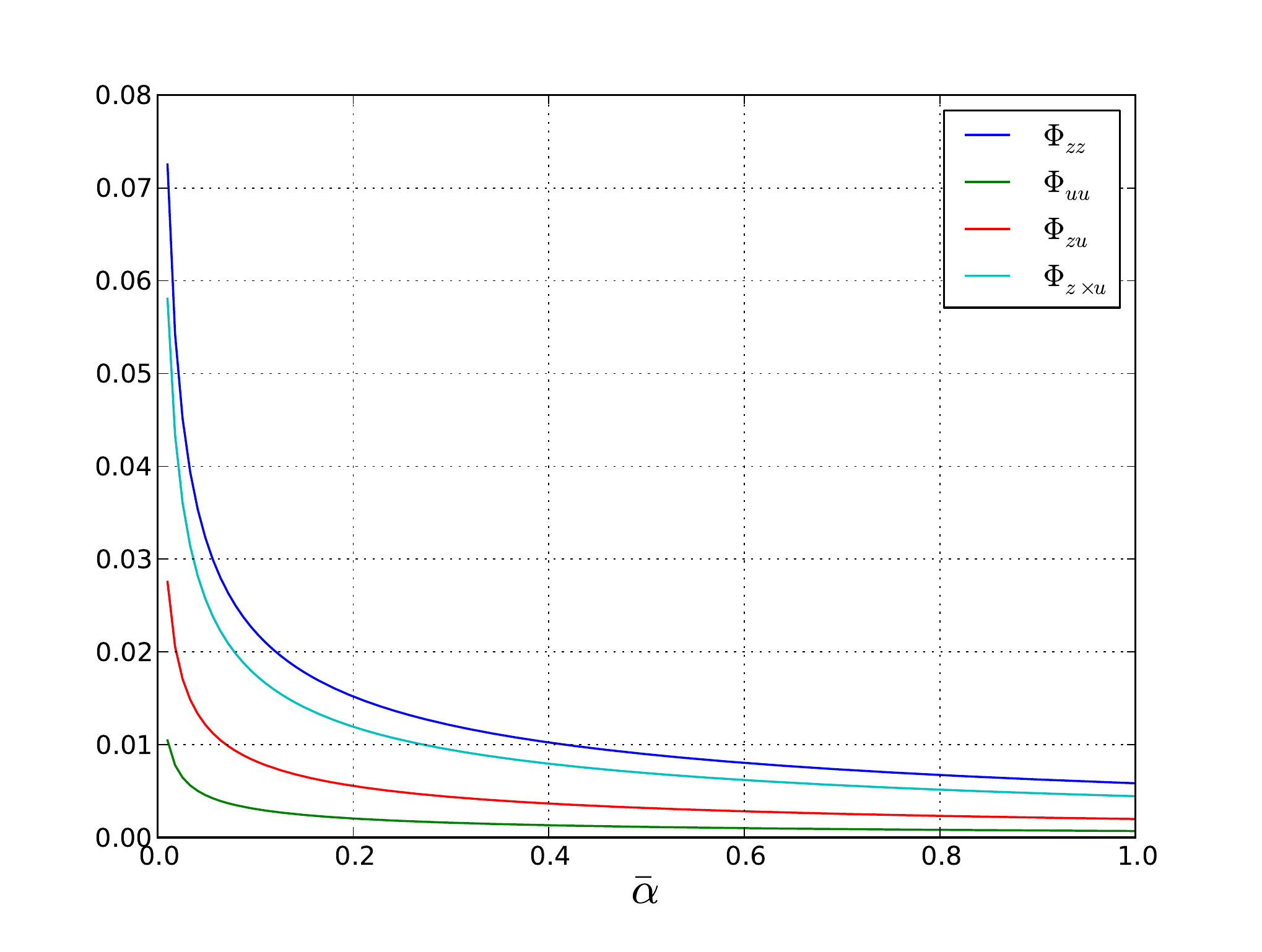}
\caption{Dependence of noise on the pinning strength $\alpha$ for $z_0=u_0=0.8\lambda$ and $\kappa=100$ where $\bar\alpha=2\alpha (4\pi\xi\lambda)^2/\phi_0^2$. The 4 curves correspond to the 4 noise components as ordered in the inset.}
\end{figure}

\begin{figure}[htb]
\includegraphics[trim = 80px 40px 10px 50px, clip,scale=0.5]{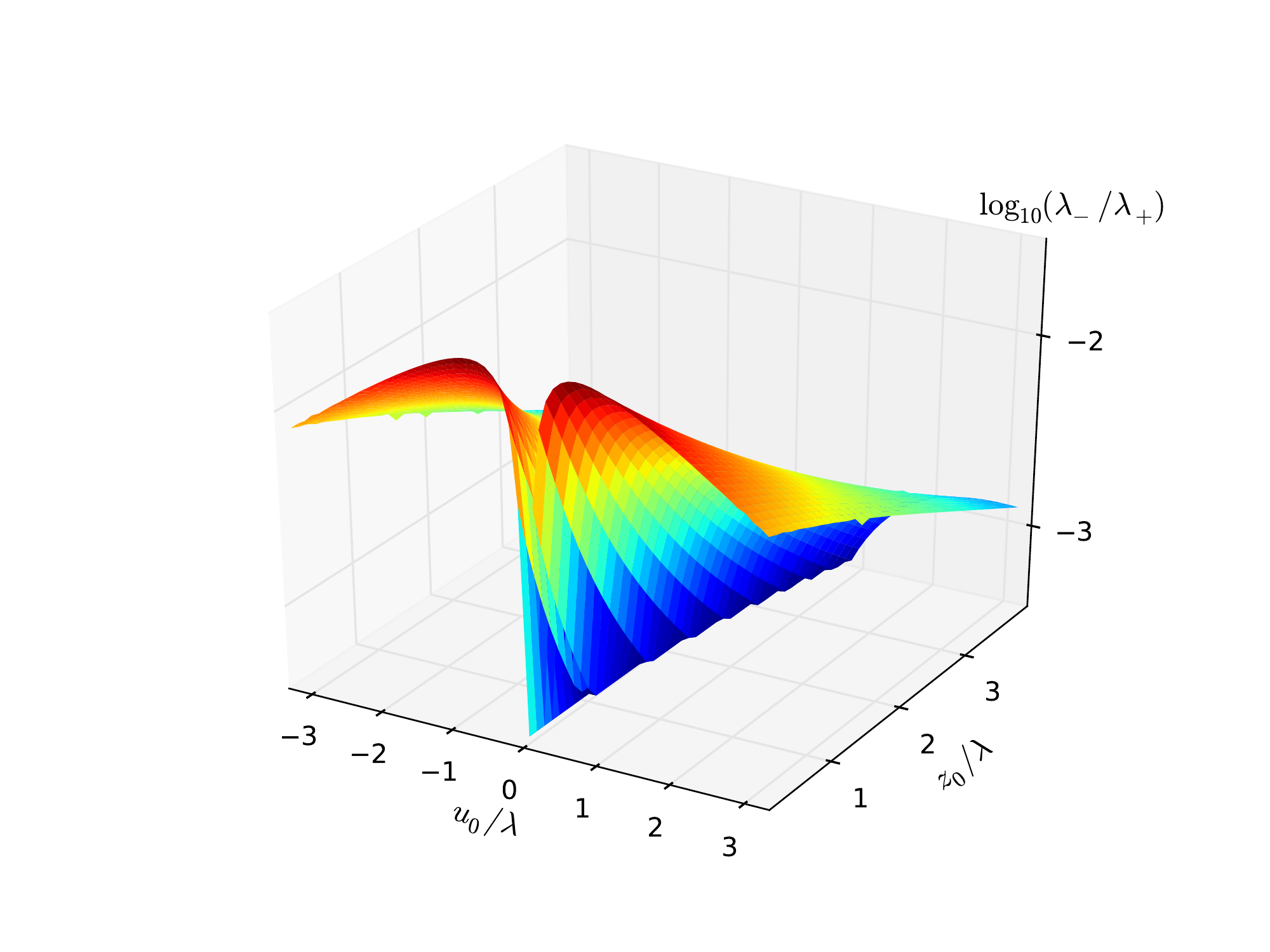}
\caption{Ratio of the two noise eigenvalues in the $z_0,u_0$ plane. Parameters are $\bar\alpha=0.1,\,
\kappa=100$. }
\end{figure}

\begin{figure}[htb]
\includegraphics[scale=0.38]{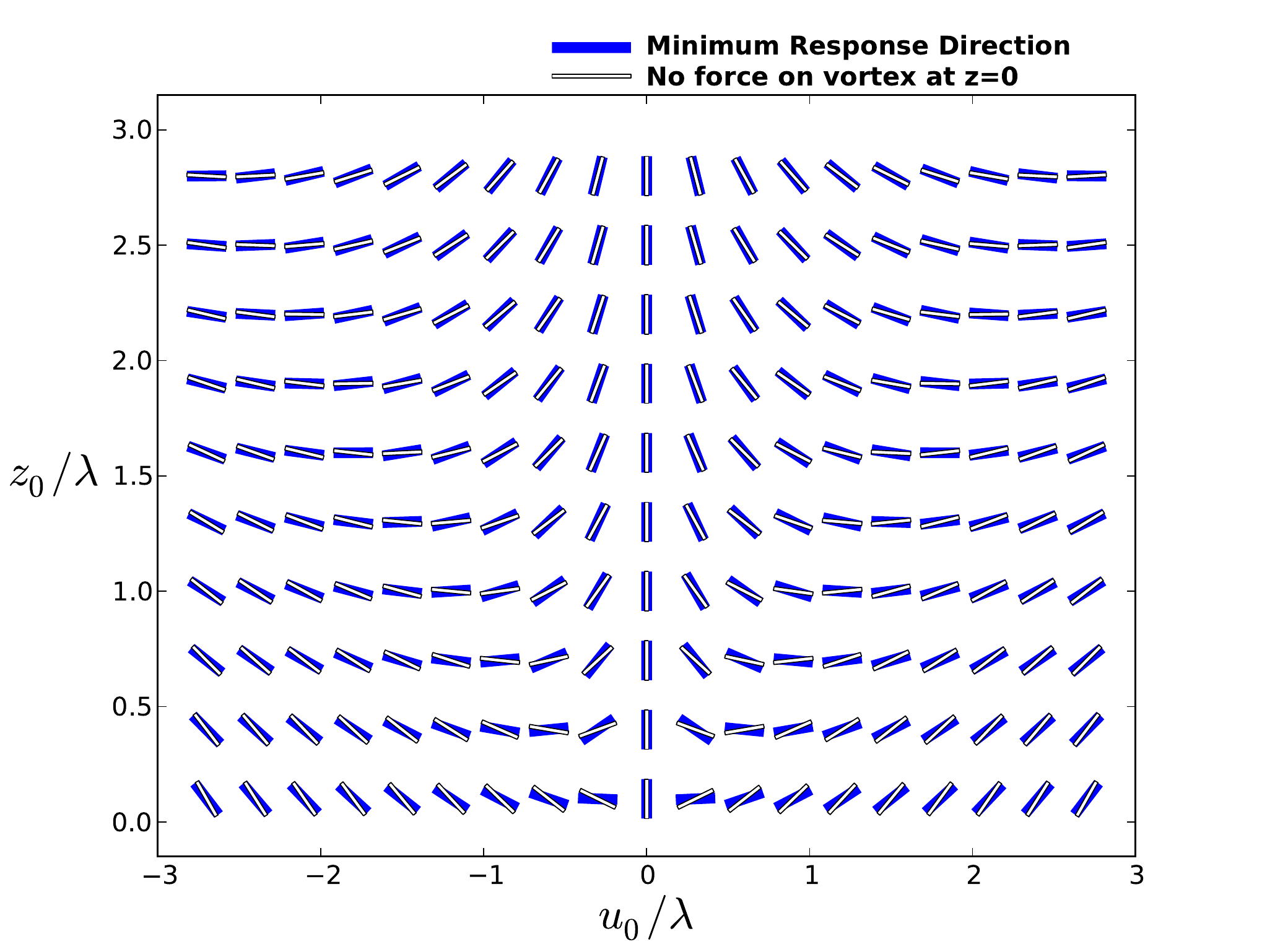}
\caption{Direction of the eigenvector with minimum magnetic fluctuations (full lines). The white insertions are the directions for which there is no force on the vortex at $z=0$. Parameters are $\bar\alpha=0.1,\,
\kappa=100$.}
\end{figure}

For not too weak pinning, $\bar\alpha\gg \ln\kappa/\kappa^2$, we have in Eq. (\ref{e21}) $f(\alpha)\rightarrow \frac{\lambda^6}{2\bar\alpha \kappa^2}\ln\kappa$, and $\Phi_{ij}\sim 1/\alpha$, hence $\im\alpha_{ij}^{(1)}\sim 1/\alpha^2$. Fig. 2 shows the $\alpha$ dependence at an intermediate scale, displaying a somewhat stronger decrease with $\alpha$.

The response in the $z-u$ plane can be diagonalized leading to Fig. 3 with the ratio of upper and lower eigenvalues. We find that the magnetic noise is reduced by at least a $10^2$  in the direction of the eigenvector with the lower eigenvalue. This direction is shown in Fig. 4 by the thick blue lines. We find that these directions correlate well with the direction of $\mm$ such that the source (\ref{e16}) vanishes at $z=0$. The latter directions are shown as the inner white lines in Fig. 4. We note that for $z_0^2+u_0^2\gg\lambda^2$ the lower eigenvalue of Eq. (\ref{e21}) actually vanishes and its eigenvector in the $(z_0,u_0)$ plane is $\sim [2u_0^2-z_0^2,3z_0u_0]$, in agreement with Fig. 4.

To appreciate the scale, we note that for a normal metal the response is (for $z<\delta$) $\im \alpha^n_{zz}=
\frac{\pi\omega}{2 z_0 c^3 \rho_n}$ hence the ratio is (similar for other components)
\beq{22}
\frac{\im\alpha^{(1)}_{z,z}}{\im\alpha^n_{z,z}} = \frac{128\pi}{\bar\alpha\ln\kappa}\Phi_{zz}(u_0,z_0)\frac{z_0}{\lambda}
\eeq
 For $z_0\approx\lambda$ and ${\bar\alpha}\approx 1$ the overall magnitude, as seen in Fig. 1, is comparable to that of a normal metal. Hence at short scales $z_0,u_0 \lesssim \lambda$ the vortex fluctuations lead to relatively low noise only in the special directions . At large distance $z_0\gg\lambda$ the noise decreases rapidly as $z_0^{-\sigma}$ with $\sigma=6-8$ (Eq. \ref{e21}) for the various components, while the normal metal's noise decreases as $1/z_0$. Hence in the range $\lambda\ll z_0\ll \delta$ the vortex noise is considerably less than that of the normal metal.

Finally we consider the case $\alpha=0$ corresponding to weak pinning $\alpha\ll \eta\omega$. This case can correspond to weak pinning from point defects, such that the persistence length of the vortex \cite{blatter} $L_c\gg L$ (and $L\gg\lambda$ for our solution to hold). In this case the vortex elasticity overcomes the pinning and the vortex remains essentially straight in equilibrium.
We define a dimensionless parameter $(\frac{4\pi\lambda^2}{\phi_0})^2\frac{\eta\omega}{2\ln\kappa}={\bar\eta}\omega$ which with the BS friction, $\kappa=10^2$ and typical parameters as above is $\approx 10^{-2}$. For small ${\bar\eta}\omega$ we find that the response has the form $\im\alpha^{(2)}_{ij} = \frac{\phi_0}{\lambda^5c\sqrt{2\omega\eta\ln\kappa}}\bar\Phi_{ij}(u_0,z_0)$ where $\bar\Phi_{ij}(u_0,z_0)$ is $\eta$ independent. In Eq. (\ref{e17}) $\eta$ provides a restoring force, hence a divergence $\sim 1/\sqrt{\eta}$ as $\eta\rightarrow 0$. Therefore, for a given $\alpha$ and as $\eta$ is reduced, in the first range $\alpha\ll\eta\omega$ the noise increases, becomes maximal at $\alpha\approx\eta\omega$, then in the regime $\alpha\gg\eta\omega$ the noise decreases as $\im\alpha^{(1)}_{ij}\sim \omega\eta/\alpha^2$ and finally vanishes at $\eta\rightarrow 0$.

Diagonalizing $\bar\Phi_{ij}$ leads to an eigenvalue $\lambda_-=0$ with an eigenvector which is very close to that of the strong pinning case in Fig. 4.
We find that $\bar\Phi_{ij}$ and $\Phi_{ij}$ have a very similar $z_0,u_0$ dependence. In particular for $z_0^2+u_0^2\gg\lambda^2$ we find that $\bar\Phi_{ij}$ is given by Eq. (\ref{e21}), except for the replacement $f(\alpha)\rightarrow \lambda^6$. Hence, comparison with the strong pinning case yields
\beq{23}
\frac{\im\alpha^{(2)}_{ij}}{\im\alpha^{(1)}_{ij}} = \frac{\bar\alpha\pi\sqrt{2}\kappa^2}{(2\bar\eta\omega)^{3/2}\ln\kappa}
\frac{\bar\Phi_{ij}(u_0,z_0)}{\Phi_{ij}(u_0,z_0)}\rightarrow \frac{\pi\bar\alpha^2 \kappa^4}{(\bar\eta \omega)^{3/2}\ln^2\kappa}
\eeq
where the limit corresponds to $z_0^2+u_0^2\gg\lambda^2$. For $\bar\alpha$ not too small this ratio is large, in particular due to the $\kappa^4$ factor, originating from $\im\alpha^{(1)}_{ij}\sim 1/\alpha^2$.

\section{conclusions}
We present here a systematic treatment for a vortex response to an external field. Our formalism provides a rigorous derivation of the boundary condition (\ref{e14}) for an arbitrary vortex shape. We apply our results to the problem of dephasing and spin flip in cold atom traps. The single vortex provides an efficient tool for trapping cold atoms \cite{shimizu,muller1,muller2}, hence the significance of evaluating the fluctuation effects.

 The single vortex in our system breaks translational symmetry, hence the reflection from the surface is not specular and off-diagonal elements appear in the response matrix. Due to these elements, there is a special
 $(z_0,u_0)$ dependent direction, for which the fluctuations are considerably reduced.

 For a trapping static field in the $\hat{\bf n}$ direction, the magnetic fluctuations $\sim\im\alpha_{ii}$ measure the fluctuations in the energy difference of two trapped magnetic sublevels with $i$ in the $\hat{\bf n}$ direction, leading to dephasing \cite{scheel,folman}. By choosing $\hat{\bf n}$ parallel to that of the minimum noise, Fig. 4, dephasing will be considerably reduced. In contrast, the sf rate depends on field fluctuations perpendicular to $\hat{\bf n}$, i.e. the  directions in $\sum_i$ of Eq. (\ref{e01}). Hence by choosing the trap direction perpendicular to those in Fig. 4 the noise will be reduced by a factor $\approx 2$.

 We find that strong pinning, e.g. as from a columnar defect, is significant for reducing magnetic noise. Furthermore, we find a strong decay at $z_0>\lambda$ of the magnetic fluctuations $\sim z_0^{-\sigma}$, with $\sigma=6-8$ for various noise components, Eq. (\ref{e21}). The regime $\lambda\ll z_0\ll\delta$ is a regime where the vortex static field ${\bf B}$  is still significant for trapping \cite{carneiro}, i.e. $B_z\sim 1/z_0^2, B_u\sim 1/z_0^3$. Hence in this regime the magnetic fluctuations are significantly reduced, allowing for efficient trapping with low dephasing and spin flip rates.

 \acknowledgments
We acknowledge highly stimulating and valuable discussions with H. R. Haakh, C. Henkel and B. Pla\c{c}ais. This research was supported by a Grant from the G.I.F., the German-Israeli Foundation for Scientific Research and Development.


\hide{
\beq{19}
\im\hh^r = \frac{4\omega\eta}{\lambda^3 \alpha c\ln\kappa}
\left(   \begin{array}{ccc}
\Phi_{zz}(u_0,z_0)&\Phi_{zu}(u_0,z_0)&0\\
\Phi_{uz}(u_0,z_0)&\Phi_{uu}(u_0,z_0)&0\\
0&0&\Phi_{z\times u_0}(u_0,z_0)
\end{array}\right)
\left(   \begin{array}{c}m_z\\ m_{u0}\\ m_{z\times u_0}\end{array}\right)
\eeq
}

\end{document}